\documentclass[iopart,amsmath,amssymb,twocolumn,showpacs]{revtex4-2}
\usepackage{graphicx}
\usepackage{dcolumn}
\usepackage{bm}
\usepackage{color}
\usepackage[paperwidth=210mm,paperheight=297mm,centering,hmargin=2cm,vmargin=2.5cm]{geometry} 

\begin{document}

\title{Numerical study of magneto-optical binding between two dipolar particles under illumination by two counter-propagating waves} 

\author{Ricardo Mart\'{\i}n Abraham-Ekeroth}
\affiliation{Instituto de F\'{\i}sica Arroyo Seco, IFAS (UNCPBA), Tandil, Argentina} 

\affiliation{CIFICEN (UNCPBA-CICPBA-CONICET), Grupo de Plasmas Densos, Pinto 399, 7000 Tandil, Argentina}
\email{mabraham@ifas.exa.unicen.edu.ar}

\begin{abstract}
	The formation of a stable magneto plasmonic dimer with THz resonances is theoretically studied for the principal directions of the system. Unlike a recent report, our work provides a complete description of the full photonic coupling for arbitrary magnetic fields as, for instance, unbalanced particle spins. As an illustration, we consider two small, n-doped InSb nanoparticles under illumination by two counter-propagating plane waves. Remarkably, when an external magnetic field exists, the symmetry in the system is broken, and a resonant radiation pressure for the dimer appears. Similarly, tunable inter-particle forces and spins are exerted on the non-reciprocal dimer. The system is also characterized when the magnetic field is absent.
Moreover, we show how the mechanical observables truly characterize the dimer since their resonance dependency contains detailed information about the system. Unlike far-field observables like absorption, mechanical magnitudes depend on the system's near-field. In addition, the nature of the particle spins is originally explained by the energy flow's behavior around the dimer.
This work constitutes a generalization of any previous approach to optical binding between small nanoparticles. It paves the way for fully controlling optical matter and nano factory designs based on surface plasmon polaritons.
\end{abstract}

\maketitle

\section{Introduction}

Optical matter (OM) consists of arrays of micro or nanoparticles that are somehow bound and controlled by light \cite{burns_optical_1990}. An OM able to self-assemble at will to develop solid technology is a long-standing goal in photonics \cite{parker_optical_2020}. The background for OM is the particle manipulation by optical forces; the first results were applied to microparticles due to the lack of technology and the presence of thermal noise for smaller systems \cite{ashkin_optical_1975,ashkin_internal_1989}. Generally, OM comprises multiple particles subjected to electromagnetic forces that come from their mutually scattered light. However, this multiple scattering phenomena can be very complex and bring about unusual effects such as “non-reciprocal” forces, torque opposite to the illumination angular momentum, and non-conservative forces \cite{albaladejo_scattering_2009,parker_optical_2020}.
Appropriate control of OM by forces and torques could lead to programmable materials for optomechanical, rheological, and biological applications. In this respect, many works studied the optical binding between nanoparticles as a primary tool to develop OM. Some approached specific combinations of optical beams like those with programmable phase \cite{han_crossover_2018,nan_creating_2022}. For example, light-induced rotation of objects holds potential for various applications such as sensing, cargo transportation, drug delivery, and micro/nanosurgery \cite{xin_optical_2020,arita_coherent_2020,blazquez-castro_genetic_2020}. Optical traps use the combination of beams as a potent characterization tool for material science and biophysics, as in Ref.~\cite{erdogan_atmospheric_2022}, which uses electrostatic focusing to obtain the mass spectrum of SARS-CoV-2 and BoHV-1 virions. However, the high intensity at the focal spot may introduce laser heating, which is an issue for bio applications \cite{nan_creating_2022}.

On the other hand, recent advances in THz technology call for new devices and materials that exhibit a non-reciprocal behavior for photonic networks and optical information processing \cite{shui_optical_2022}. Non-reciprocal devices are a crucial component of modern communication technology. They are nowadays required for miniaturized electronic and photonic devices \cite{nature_directions_2020}. One way to create optical non-reciprocity at the THz range is using magneto-optical (MO) systems like graphene, hexaferrites, and semiconductors \cite{kimel_2022_2022}. For instance, Ref.~\cite{chochol_magneto-optical_2016} presented MO measurement of several samples of InSb with different carriers and carrier concentrations for low external magnetic field and room temperatures. With these advantages, dimers and trimers of InSb particles have been studied to enhance THz spectroscopy by forming electric and magnetic hotspots in the gap between them \cite{bakker_magnetic_2015,sadrara_electric_2019}. Anisotropic materials like InSb in OM would make it strongly dependent on the beam combinations, allowing for countless possibilities \cite{roichman_optical_2008}.
Recently, several efforts have focused on MO nanoparticle systems to shape OM and optical traps with a reasonable degree of control and accuracy, besides other relevant applications \cite{emile_observation_1992,hu_trapping_2013,abraham_ekeroth_thermal_2017-2,edelstein_magneto-optical_2019-1,edelstein_circular_2022}. on a broader sense, magnetoplasmonics relates the plasmonic behavior of nanoparticles with the presence of external magnetic fields. The modulation and tunability of plasmonic resonances offered by magnetoplasmonics results auspicious for ultra-sensitive sensors and active plasmonic devices \cite{lacroix_active_2012}. In particular, the formation of stable optical binding between two small magnetoplasmonic particles has been lately studied \cite{edelstein_magneto-optical_2021}. Equilibrium binding distances were predicted and found tunable by the incoming wave's polarization state and the magnetic field's magnitude. However, the model developed in that report is valid only for relatively small magnetic field values. Moreover, it predicts stable dimers only using alternating magnetic static fields and polarization angles to remove azimuthal, unbalanced forces. More importantly, this work needs to discuss possible rotations of the particles due to angular momentum transfer in the multiple scattering scheme. 

In this paper, we study the formation of stable MO dimers for small nanoparticles in a complete framework involving all the possible optomechanical inductions. The dimer's isotropic and anisotropic responses are assessed as a base for OM designs, i.e., under the presence/absence of an external magnetic field. This field can be of arbitrary magnitude in our model. The illumination consists of two counter-propagating plane waves with circular polarization, which simulates a simple optical trap in the vacuum. We found several possibilities to create stable dimers even when the magnetic field is off. The beams do not exert net forces for reciprocal dimers but may exert torques on them.
On the contrary, there is a net radiation pressure and spin for the whole system when the static field is on, allowing complete control of the system's movement. The results will enable one to infer that the mechanical variables can be used as near-field observables to explore the content of unknown samples. Conversely, they can be used to accurately control the dimer's creation/destruction and its mobility. Finally, the spins predicted are explained in terms of the energy flows around the dimer, which constitutes a novel scattering-force effect for interacting particle arrays.

\section{Model}

\begin{figure}
	\begin{centering}
		\includegraphics[width=8.5cm,keepaspectratio]{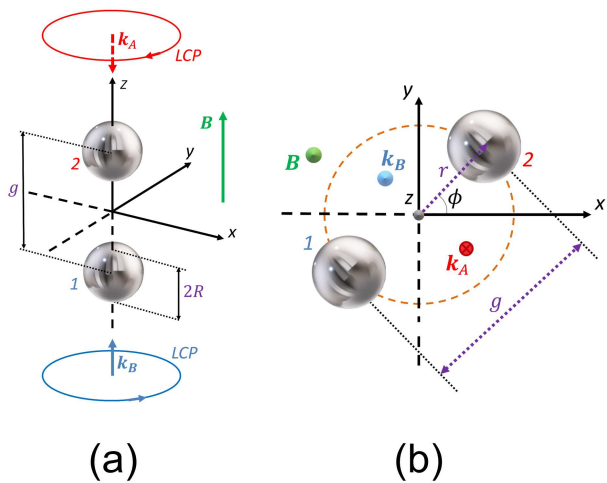}
		\par\end{centering}
	\caption{\label{fig:1_GralConfig}(Color online) Dimer configurations and incident waves treated in this work. Two counter-propagating waves with left circular polarization illuminate the magneto-optical dimer. The leading example consists of two n-doped InSb particles separated by a gap of a particle's diameter, $g=2R$. (a) Parallel [(b) perperdicular] configuration. The static magnetic field $B$ is parallel to $+z$ direction (green arrow). In (b), $\phi$ is the azimuthal angle of the dimer's position.}
\end{figure}

In the following, we assume two equal particles of the same non-reciprocal material immersed in the vacuum. Then the method of discrete dipoles (MDD) \cite{de_sousa_magneto-optical_2016} simplifies considerably to
\begin{align}
	\label{eq-dipoles}
	\mathbf p_{1} =  \epsilon_0\hat \alpha \mathbf E_{0,1} + k^2_0 \hat \alpha \hat G \mathbf p_{2}, \\
	\mathbf p_{2} =  \epsilon_0\hat \alpha \mathbf E_{0,2} + k^2_0 \hat \alpha \hat G \mathbf p_{1},
\end{align}
where $\hat \alpha$ is the polarizability tensor representing the particles. The following definition automatically includes the radiative corrections necessary to fulfill the optical theorem \cite{ekeroth_optical_2019-1}
\begin{equation}
	\label{eq-alpha}
	\hat \alpha = \left(\hat \alpha_0^{-1} - \frac{ik^3_0\hat I}{6\pi}\right)^{-1}
\end{equation}
where $\hat \alpha_0$ is the so-called quasistatic polarizability, which can be given by
\begin{equation}
	\label{eq-alpha0}
	\hat \alpha_{0}^{-1} = \frac{1}{V} \left( \hat L + [\hat \epsilon_r - \hat I]^{-1} 
	\right)
\end{equation}
being $V$ the particles' volume, $\hat L=\hat I/3$ is the electrostatic depolarization tensor specified for spheres or cubes, and $\hat \epsilon_r$ is the relative dielectric tensor. The system of equations \ref{eq-dipoles} can be solved straightforwardly, leading to
\begin{align}
	\label{eq-dipoles-solved}
	\mathbf p_{1} = \epsilon_0 \hat F \left(\mathbf E_{0,1} + k^2_0 \hat \alpha \hat G \hat \alpha \mathbf E_{0,2}\right), \\
	\mathbf p_{2} = \epsilon_0 \hat F \left(\mathbf E_{0,2} + k^2_0 \hat \alpha \hat G \hat \alpha \mathbf E_{0,1}\right),
\end{align}
where we define $\hat F= \left(\hat \alpha - k^4_0 \hat G \hat \alpha \hat G\right)^{-1}$. In this work, a counter-propagating configuration is assumed as a superposition of two left-handed circularly polarized (LCP) plane waves with the same intensity $I_0$ \cite{cameron_optical_2014,edelstein_magneto-optical_2019-1}, see Fig.~\ref{fig:1_GralConfig}, namely,
\begin{equation}
	\label{eq-E0}
	\mathbf E_{0} = \frac{E_{0}}{\sqrt{2}}\left[\left(\mathbf{\check{x}}+i\mathbf{\check{y}}\right)e^{ik_0z}+\left(\mathbf{\check{x}}-i\mathbf{\check{y}}\right)e^{-ik_0z}\right].
\end{equation}
This field is used in Eq.~\ref{eq-dipoles-solved} to calculate the incident field at the particles' positions $\mathbf E_{0,1}$ and $\mathbf E_{0,2}$.

The absorption cross section of the system can be calculated once the dipole moments are known by
\begin{align}
	& \sigma_{abs}=\frac{k_0}{\epsilon_0 w^{tot}_E}Im\left\{\mathbf{p}_1\cdot\left(\hat \alpha_0^{-1}\mathbf{p}_1\right)^* + \mathbf{p}_2\cdot\left(\hat \alpha_0^{-1}\mathbf{p}_2\right)^*\right\}	
\end{align}
where $w^{tot}_E=\epsilon_0|\mathbf{E}_0|^2$. The $i$-component of the forces exerted on each particle can be obtained from the time-averaged force within the Rayleigh approximation \cite{chaumet_time-averaged_2000}. This is 
\begin{eqnarray}
	\label{eq-FcDDA}
	F_{1,i}& = &\frac{1}{2}Re\{\mathbf{p}^t_1[\partial_i\mathbf{E}^{*}(\mathbf{r},\omega)|_{\mathbf{r}=\mathbf{r}_1}\} \\
	F_{2,i}& = &\frac{1}{2}Re\{\mathbf{p}^t_2[\partial_i\mathbf{E}^{*}(\mathbf{r},\omega)|_{\mathbf{r}=\mathbf{r}_2}\}
\end{eqnarray}
where the derivatives of the total field $\partial_i\mathbf{E}(\mathbf{r},\omega)|_{\mathbf{r}=\mathbf{r}_n}$ at the dipoles' positions $\mathbf{r}_n$, $n=\{1,2\}$ can be obtained from \cite{chaumet_coupled_2007}:
\begin{align}
	\label{eq-derivsEr}
	& \partial_i\mathbf{E}(\mathbf{r},\omega)|_{\mathbf{r}=\mathbf{r}_1}=\partial_i\mathbf{E}_0(\mathbf{r},\omega)|_{\mathbf{r}=\mathbf{r}_1}+ \nonumber \\ 
	& +\frac{k^2_0}{\epsilon_0}(\partial_i\mathbf{G}(\mathbf{r},\mathbf{r}_2))_{\mathbf{r}=\mathbf{r}_1}\mathbf{p}_2]\}  \\
	&\partial_i\mathbf{E}(\mathbf{r},\omega)|_{\mathbf{r}=\mathbf{r}_2}=\partial_i\mathbf{E}_0(\mathbf{r},\omega)|_{\mathbf{r}=\mathbf{r}_2} + \nonumber \\ 
	& +\frac{k^2_0}{\epsilon_0}(\partial_i\mathbf{G}(\mathbf{r}_1,\mathbf{r}))_{\mathbf{r}=\mathbf{r}_2}\mathbf{p}_1]\} 
\end{align}
The total force exerted on the dimer results from adding the force components for each particle, namely, $F_{tot,i} = F_{1,i}+F_{2,i}$. In particular, the net radiation pressure for the dimer under the illumination given by Eq.~\ref{eq-E0} is defined by taking $i=3$, or the $z$ components, as
\begin{equation}
	\label{eq-ESys}
	F_{tot,z} = F_{1,z}+F_{2,z}
\end{equation}
Another useful mechanical variable is the binding force, which in the present case is defined as
\begin{equation}
	\label{eq-BindingF}
	\Delta = \left(\mathbf{F}_1 - \mathbf{F}_2\right)\cdot \check{n}
\end{equation}
where $\mathbf{\check{n}}=\frac{\mathbf{r}_2-\mathbf{r}_1}{|\mathbf{r}_2-\mathbf{r}_1|}$ is the dimer's versor. The optical torques can also be calculated, as given in Ref.~\cite{chaumet_electromagnetic_2009}:
\begin{align}
	\label{eq-Tqs1}
	& \mathbf{N}_{spin,1} = \frac{1}{2\epsilon_0}Re\left\{\mathbf{p}_{1} \times \left[\left(\hat \alpha_0^{-1}\right)^*\mathbf{p}_1^*\right]\right\} \\
	& \mathbf{N}_{orb,1} = \mathbf{r}_{1} \times \mathbf{F}_{1} \\
	& \mathbf{N}_{1} = \mathbf{N}_{spin,1} + \mathbf{N}_{orb,1}
\end{align}
\begin{align}
	\label{eq-Tqs2}
	& \mathbf{N}_{spin,2} = \frac{1}{2\epsilon_0}Re\left\{\mathbf{p}_{2} \times \left[\left(\hat \alpha_0^{-1}\right)^*\mathbf{p}_2^*\right]\right\} \\
	& \mathbf{N}_{orb,2} = \mathbf{r}_{2} \times \mathbf{F}_{2} \\
	& \mathbf{N}_{2} = \mathbf{N}_{spin,2} + \mathbf{N}_{orb,2}
\end{align}
The definitions of the orbital and spin torques were discussed previously in Refs.~\cite{nieto-vesperinas_optical_2015,nieto-vesperinas_optical_2015-1,chaumet_electromagnetic_2009}, among others. The spin torques are always defined with respect to the centers of the particles. Otherwise, the reference system is located at the dimer's center of mass, and orbital torques are set. Thus, the total torque exerted on the dimer is
\begin{align}
	\label{eq-TqSys}
	& \mathbf{N}_{orb} = \mathbf{N}_{orb,1} + \mathbf{N}_{orb,2} \\
	& \mathbf{N}_{spin} = \mathbf{N}_{spin,1} + \mathbf{N}_{spin,2} \\
	& \mathbf{N}_{tot} = \mathbf{N}_{1} + \mathbf{N}_{2}=\mathbf{N}_{orb}+\mathbf{N}_{spin}
\end{align}
In particular, this study simulates nanoparticles made of n-doped Indium antimonide (n-InSb) \cite{palik_coupled_1976}. Indium antimonide (InSb) is one example of the most widely studied polar semiconductors for magnetoplasmonic applications because it can be easily doped for sizable magnetic-induced effects \cite{moncada-villa_magnetic_2015,chochol_magneto-optical_2016}. As reviewed in Refs.~\cite{palik_coupled_1976,moncada-villa_magnetic_2015}, n-InSb is an exciting material that has two kinds of surface resonances in the absence of static field, namely, the phonon polariton (SPhP, higher-energy) and the plasmon polariton (SPP, lower-energy). Its model properties were described on Refs.~\citep
{moncada-villa_magnetic_2015,abraham_ekeroth_thermal_2017-2,edelstein_magneto-optical_2019-1}, among others. Since we are interested in the near-field interactions between the particles, the study focuses on an example for which the interparticle's gap equals one particle diameter ($g=2R$); see Fig.~\ref{fig:1_GralConfig}. We add complementary examples for other values of the gap in the Supplementary Material (SM).

\section{Results}

In this section, all the optical variables were scaled by the proper factors to make them adimensional. The following characteristic magnitudes, namely,  $w^{tot}_E$, $A_p=\pi R^2$, $V_p=\frac{4}{3}\pi R^3$, and $V_{int}=\frac{4}{3}\pi \left(|\mathbf{r}_2-\mathbf{r}_1|\right)^3$ redefine the variables as $Q_{abs}=\frac{\sigma_{abs}}{2A_p}$ for the absorption efficiency, $F_{rad}=\frac{F_{tot,z}}{w^{tot}_E A_p}$ and $\Delta'=\frac{\Delta}{w^{tot}_E A_p}$ for the radiation pressure and the binding forces, and  $\mathbf{N}_{spin}'=\frac{\mathbf{N}_{spin}}{w^{tot}_E V_p}$ for the spin torque. The variable $\mathbf{N}_{orb}'=\frac{\mathbf{N}_{orb}}{w^{tot}_E V_{int}}$ for the orbital torque is only shown by an example in the SM since it gave negligible results unless the gap is minimal, see Figs.~S1 and S2 for details. We calculate the scaled Poynting vector as $\mathbf S = \frac{1}{2 I_0}Re\{\mathbf E \times \mathbf H^* \}$ where the magnetic field $\mathbf H$ comes from an MDD equation similar to that for $\mathbf E$ \cite{novotnyPrinciplesNanoOptics2006}. The curl of $\mathbf S$ is calculated using an appropriate tridimensional mesh around the system's near-field.

\subsection{Parallel Illumination.} 

\begin{figure*}
	\centering
	\includegraphics[width=160mm]{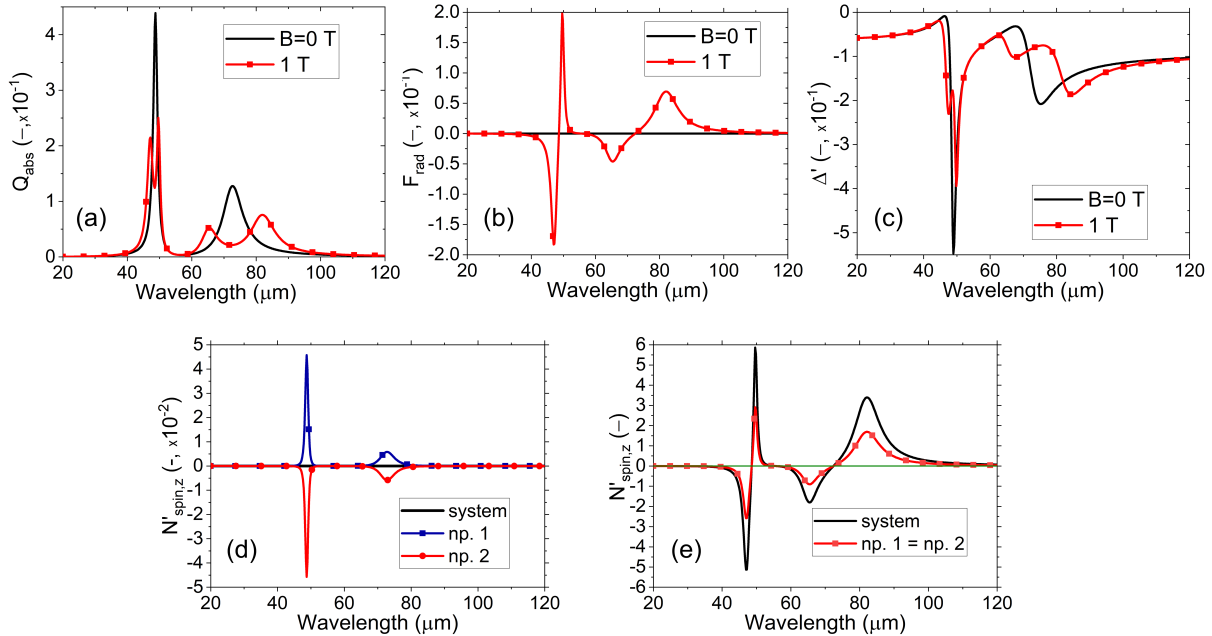}
	\caption{\label{fig:2_Parallel_Opt_prop}(Color online) Optical properties for parallel configuration. (a) Absorption efficiency, which is independent of the dimer's orientation. (b) Radiation pressure (total force along $z$). (c) Binding force. Black line [red line with squares] for magnetic field $B=0$ [$B=1$] T. (d) [(e)] Spin torques for $B=0$ [$B=1$] T. In (d), the net spin torque is zero for all wavelengths.}
\end{figure*}

Fig.~\ref{fig:2_Parallel_Opt_prop} shows the spectral results for parallel illumination when the magnetic field is off ($B=0$, black line) and on ($B=1$ T, red line with squares). The absorption efficiencies, Fig.~\ref{fig:2_Parallel_Opt_prop}a, result independent of the direction of the dimer so that the same spectra remain for any other illumination configuration. The low-energy resonance (around $73 \mu$m) corresponds to an SPP, while the high-energy resonance ($48.7 \mu$m) corresponds to an SPhP \cite{palik_coupled_1976,moncada-villa_magnetic_2015}. Making use of the Plasmon Hybridization Model (PHM) for two dipolar particles, both kinds of surface modes show as bright antibonding modes for transverse electric fields according to the configuration shown in Fig.~\ref{fig:1_GralConfig}a, see Ref.~\cite{nordlander_plasmon_2004} for details.

When $B$ is on, each isotropic surface mode splits into two modes due to degeneracy removal. The absorption is the only far-field observable shown in this work since negligible scattering occurs for small systems \cite{bohren_absorption_1998}. As it is illumination-independent, the spectra obtained remain invariant for all illumination directions concerning the dimer's axis. Thus, the only available far-field observable is neither adequate to study the interactions occurring in the dimer nor valuable to predict the dimer's dynamics. In Fig.~\ref{fig:2_Parallel_Opt_prop}b, there is no net radiation pressure when $B$ is off due to the high symmetry of both the system and incident field. On the other hand, there is a resonant pressure for the MO dimer when $B$ is on, revealing the magnetoplasmonic resonances and directing the dimer upwards or downwards according to their energy. As a result of the interaction, the radiation pressure identifies the modes by the sign of the force. Fig.~\ref{fig:2_Parallel_Opt_prop}c shows that the binding force leads to repulsion between the particles for both cases, $B=0$ and $B=1$ T. In other words, there cannot be a stable dimer under this parallel configuration. The response is still resonant but less sensitive than the radiation pressure. Remarkably, the results agree with the interpretation given by the PHM.

Notably, although we are dealing with a dimer system, our results agree with those reported in \cite{edelstein_magneto-optical_2019-1} for a single particle under the same illumination. In general, the absorption efficiency $Q_{abs}$ behaves like $Re\{\alpha_{11}\}$, both the radiation pressure $F_{rad}$ and the spin $N'_{spin,z}$ on $z$ behave like $Im\{\alpha_{12}\}$, and the binding force $\Delta '$ resembles $-Re\{\alpha_{11}\}$, being $\alpha_{ij}$ the cartesian components of the polarizability tensor $\hat \alpha$. In the case of the spins in Fig.~\ref{fig:2_Parallel_Opt_prop}d, these behave like $\pm Im\{ \alpha_{33}\}$ for each particle respectively (polarizabilities not shown here). These functional dependencies are due exclusively to the type of illumination; otherwise, other $\alpha_{ij}$-terms would appear in the spectral variables \cite{edelstein_magneto-optical_2019-1}.

Following angular momentum's conservation, Fig.~\ref{fig:2_Parallel_Opt_prop}d shows that the net spin for the system is zero when $B$ is off (black line). To put it another way, the spins for each particle are equal and opposite, showing the resonant modes for the isotropic case (see red and blue lines with symbols). When $B$ is on, however (Fig.~\ref{fig:2_Parallel_Opt_prop}e), there is a net spin for the system, black line, which is twice the spin for each particle (red line with squares). The spin resonates sensitively with the dimer's modes, quite like the radiation pressure. Consequently, the spins become much stronger than those for $B$ off, compare the scales of Fig.~\ref{fig:2_Parallel_Opt_prop}d and e. Thus, the radiation pressure and spins constitute the most sensitive observables in the near field, giving a common spectral shape (compare spectra in Figs.~\ref{fig:2_Parallel_Opt_prop}b and e). 

\subsection{Perpendicular Illumination.} 

\begin{figure}
	\begin{centering}
		\includegraphics[width=8cm,keepaspectratio]{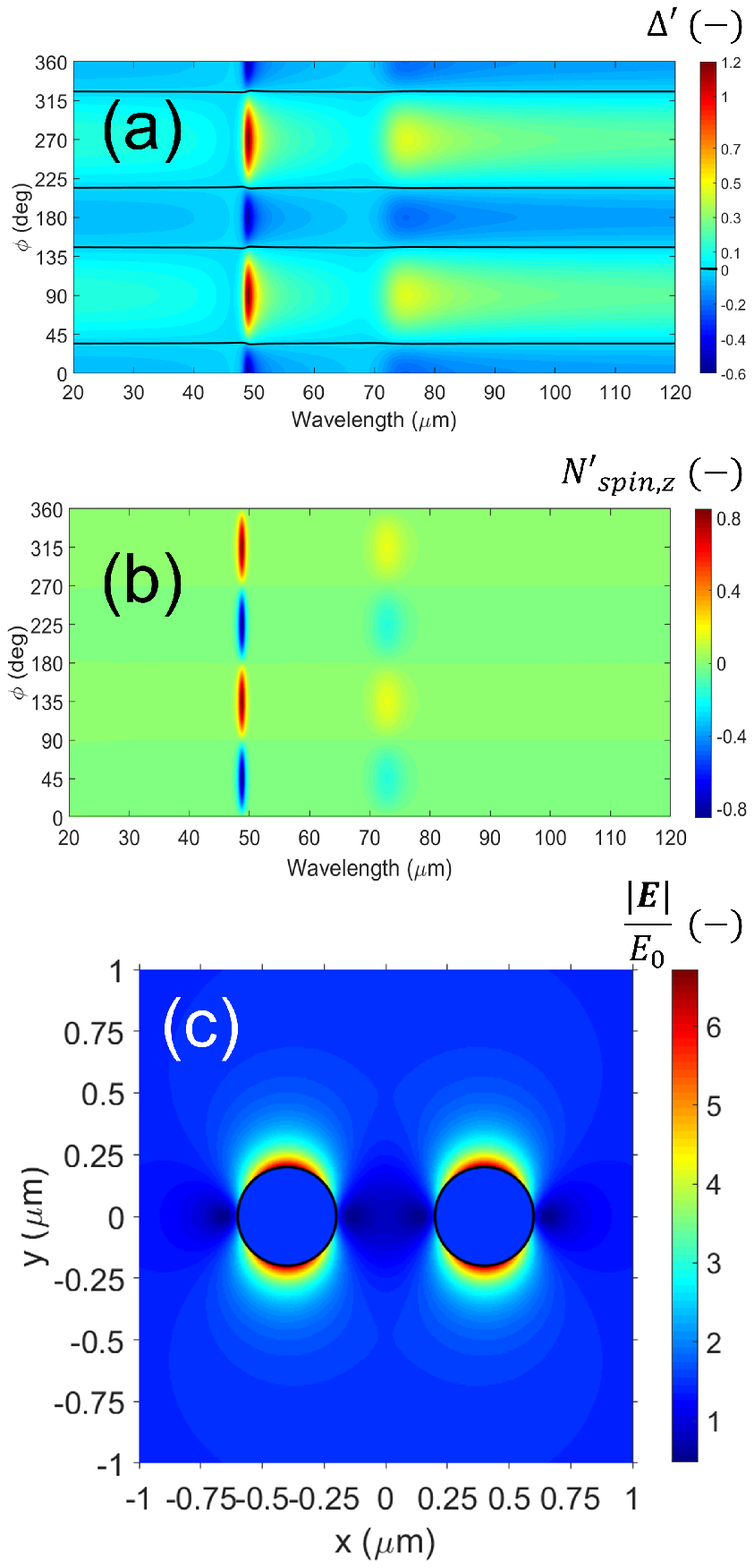}
		\par\end{centering}
	\caption{\label{fig:3_Perp_B0T}(Color online) Near-field observables for perpendicular configuration in the absence of magnetic field, $B=0$ T. (a-b) Maps of the mechanical variables as a function of wavelength and dimer's azimuthal angle. (a) Binding force. The black lines correspond to zero force. (b) Spin torque for the system. In this case, the net radiation pressure is zero for all wavelengths (not shown). (c) Distribution of electric field around the dimer for $z=0$ at the resonance wavelength $48.85 \mu$m for $\phi=0$.}
\end{figure}

\begin{figure}
	\begin{centering}
		\includegraphics[width=8.5cm,keepaspectratio]{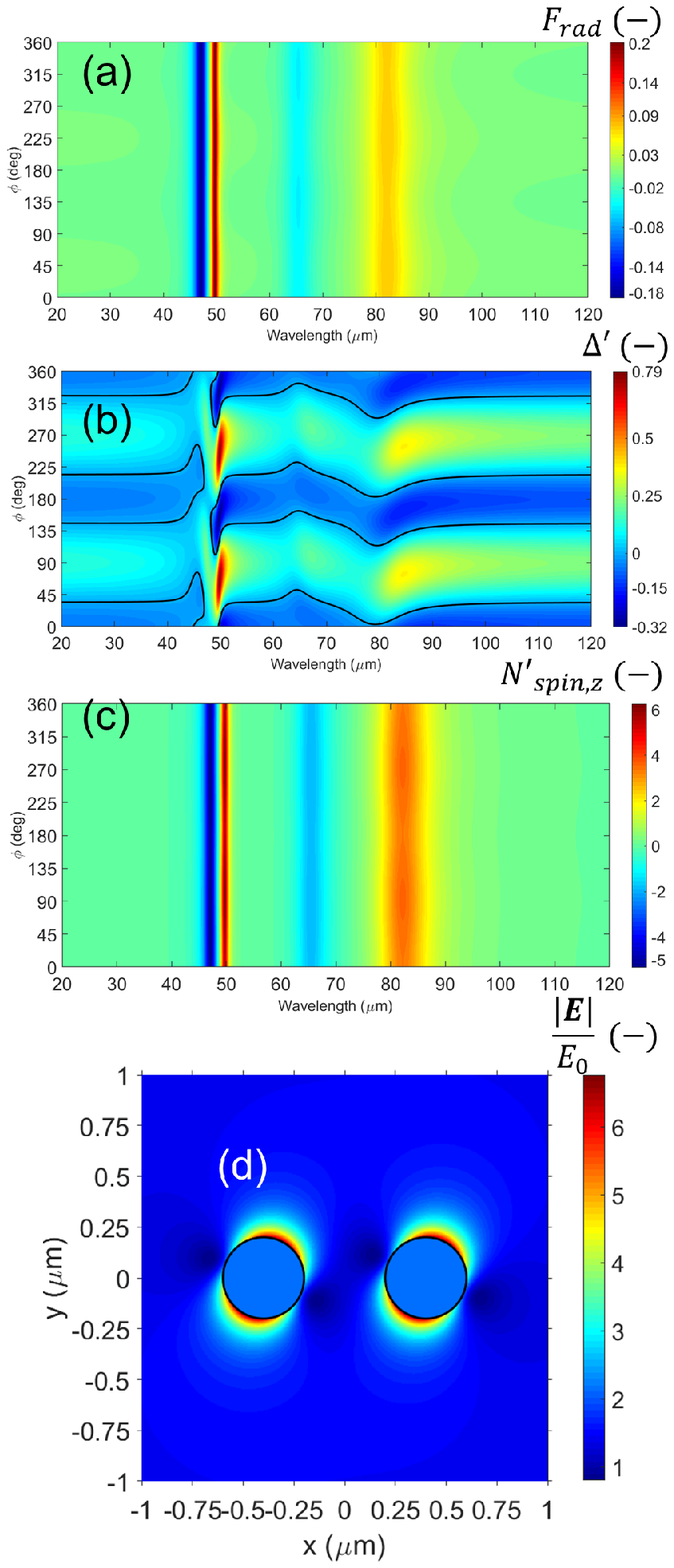}
		\par\end{centering}
	\caption{\label{fig:4_Perp_B1T}(Color online) Near-field observables for perpendicular configuration and $B=1$ T. (a-c) Maps of the mechanical variables as a function of wavelength and dimer's azimuthal angle. (a) Radiation pressure for the system. (b) Binding force. The black lines correspond to zero force. (c) Spin torque for the system. (d) Distribution of electric field around the dimer for $z=0$ at the resonant wavelength $49.85 \mu$m for $\phi=0$.}
\end{figure}

Now we vary the dimer's azimuthal angle since a dependency on the net polarization is expected. Figs.~\ref{fig:3_Perp_B0T} and \ref{fig:4_Perp_B1T} show maps as a function of the incident wavelength and azimuthal angle for $B=0$ and $B=1$ T, respectively. Similarly to that found in the previous subsection, there is no net radiation pressure when $B$ is off (not shown). Yet this time, a different behavior is found for the binding force, Fig.~\ref{fig:3_Perp_B0T}a. The system offers a resonant spectral response but depends on the angle $\phi$. Maxima [minima] of binding are found around $\phi=90,270$ [$\phi=0,180$] deg, meaning inter-particle attraction [repulsion]. This fact also defines stable positions for the dimer around the strongest optical resonance, namely the SPhP at $48.7\mu$m, and around $\phi=34,145.6,214.5,325.7$ deg for all wavelengths when $B$ is off (follow the black lines). A similar situation is found for the second resonant wavelength $\approx 72.6 \mu$m (SPP), where the variations are less pronounced. Regarding the spin, Fig.~\ref{fig:3_Perp_B0T}b shows a remaining behavior for the whole system, which is resonant with the surface modes and coordinated with the binding phenomenon. Remarkably, the spin gets its extremals (maxima or minima) when the dimer reaches its stable positions; namely, neither attraction nor repulsion, compare Figs.~\ref{fig:3_Perp_B0T}a and b. As mentioned above, the most sensitive resonance corresponds to the excitation of the SPhP. 

Noteworthy, our results are consistent with the interpretation of the PHM for isotropic, dipolar particles \cite{nordlander_plasmon_2004}. In particular, each value $\phi=n\pi$ [$\phi=(n+1/2)\pi$] rad with $n \in \mathbb{Z}$, the binding force shows repulsion [attraction] for both types of resonances, namely, the SPhP and SPP, see Fig.~\ref{fig:3_Perp_B0T}a. This outcome is due to the net polarization; the electric field is along $\mathbf{\check{y}}$, see the map for $\phi=0$ at the SPhP wavelength in Fig.~\ref{fig:3_Perp_B0T}c. Thus, $\phi=0$ corresponds to a transverse electric field compared with the dimer's direction, meaning an antibonding bright mode in the context of the PHM. Differently, $\phi=90$ deg corresponds to a parallel electric field compared with the dimer's axis, meaning a bright bonding mode in the PHM (map not shown).
In Fig.~\ref{fig:4_Perp_B1T}a, there is a remaining radiation pressure for the whole system due to the symmetry breaking that appears only at the resonances' locations. This spectrum results invariant with $\phi$ and follows the same resonances as in absorption in Fig.~\ref{fig:2_Parallel_Opt_prop}a when $B$ is on (red line with squares). Thus the presence of a static magnetic field induces the dimer to move forward or backward in the illumination's direction when the incident energy is that of a surface mode. Likewise, the system shows a resonant binding (Fig.~\ref{fig:4_Perp_B1T}b). As in Fig.~\ref{fig:3_Perp_B0T}a, the black lines follow the values of zero force. Note that the map strongly distorts by the presence of the resonances when $B$ is on, making the dynamics more complex and even reducing the extremals of the binding force. However, the possibility to obtain stable binding enhances around the SPhP due to the overlapping of MO modes, which means more degree of control in the dimer's creation and stability. 

In Fig.~\ref{fig:4_Perp_B1T}c, the system's spin follows a trend similar to that for the radiation pressure in Fig.~\ref{fig:4_Perp_B1T}a. This behavior is quite different from that found for $B=0$. Note that spin is enhanced when $B$ is on; the colorbar limits show values $\approx 6.2-7.5$ times higher than those for $B$ off, compare Figs.~\ref{fig:3_Perp_B0T}b and \ref{fig:4_Perp_B1T}c. As a result, the MO system could be readily identified in an experiment by observing the dimer's dynamics at the resonance wavelengths.

Fig.~\ref{fig:4_Perp_B1T}d shows the electric field around the dimer's plane $z=0$ for the SPhP found at $49.85 \mu$m. This wavelength corresponds to the most robust resonance when $B$ is on. The rest of the configuration is equal to that given in Fig.~\ref{fig:3_Perp_B0T}c. The field hot spots are leaned on the right $\approx 65$ deg from the $x$ axis by the MO effect.

Up to this point, we have explored a few examples of MO dimers to approach the idea of controlling the particle dynamics and "photonic molecule" stability \cite{hong_self-assembled_2013} in the presence/absence of a static magnetic field $\mathbf B$. 

Below, we discuss the behavior of the dynamic observables in terms of the information contained in the Poynting field. The reader is reminded that the particles' photonic interaction matches a multiple-scattering framework \cite{mishchenko_multiple_2006,conoir_multiple_2007}. The near fields involve the evanescent waves, which play a crucial role in the particles' interaction for surface modes and close particles. This phenomenon can be seen through the energy flows because they may have all the information of the near fields $\mathbf{E}$ and $\mathbf{H}$. Generally, the magnitudes obtained from far-field calculations lose some of the information about the system \cite{novotnyPrinciplesNanoOptics2006}. 

\subsection{Nature of the spins through an examination of the Poynting fields} 

We explore the spins exerted on the system by showing a few calculations of the Poynting field around the dimer for perpendicular configuration. The parallel configuration is less interesting since it would only lead to repulsion states without dimer formation for any gap under both cases $B=0$ and $1$ T, see Fig.~\ref{fig:2_Parallel_Opt_prop}c for the example $g=2R$. More clarifications can be found in the SM.

\begin{figure*}
	\centering
	\includegraphics[width=150mm]{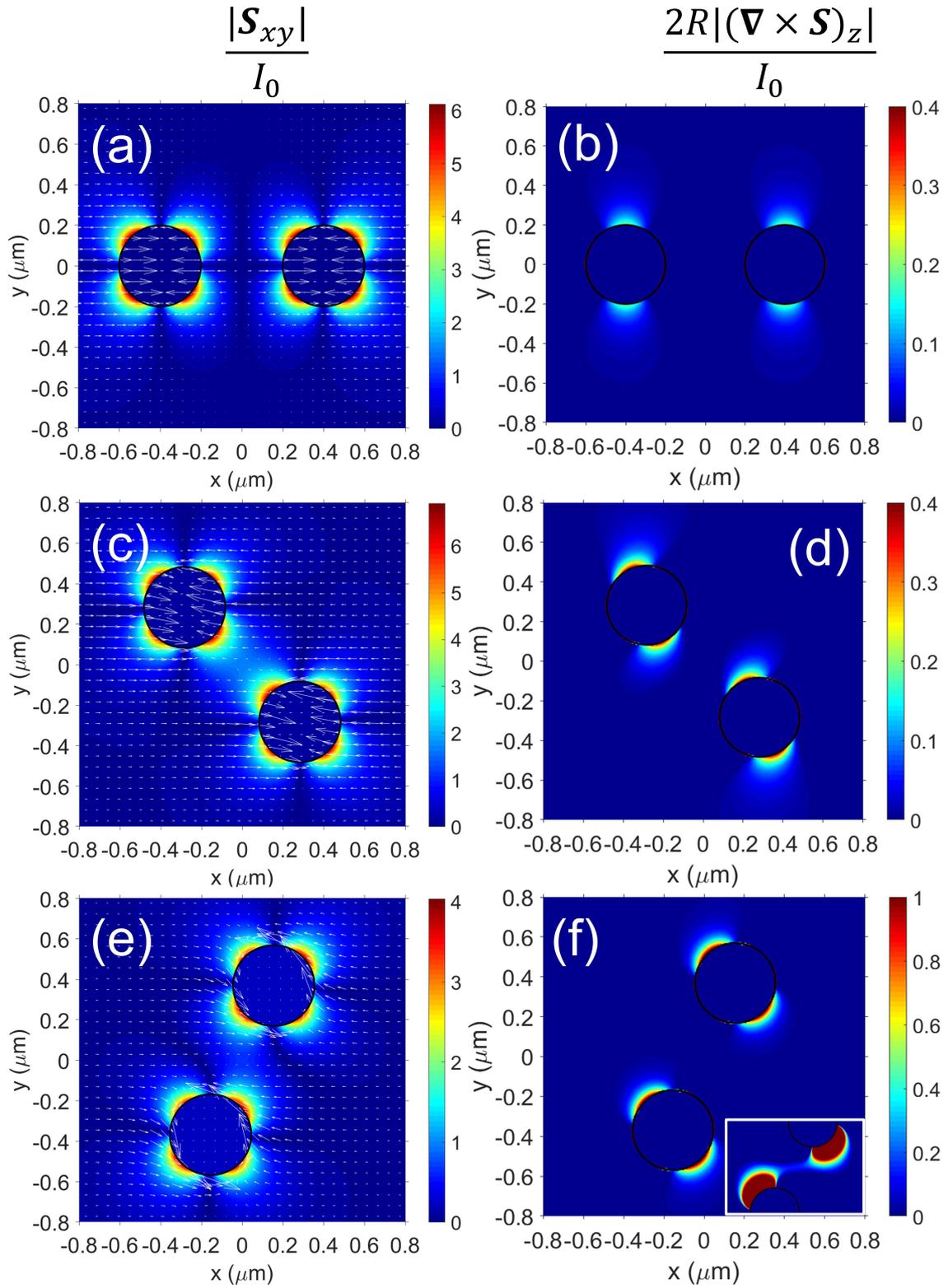}
	\caption{\label{fig:5_Perp_NF_Poynting} Scaled energy flows around the dimer for $z=0$ under perpendicular configuration and at the strongest SPhP. The color scale corresponds to the magnitude; the white arrows show the Poynting flow ($2\times$ their original size). Left [right] column for the [z-component of the curl of the] Poynting field. (a-d) [(e-f)] Examples for $B=0$ [$B=1$] T; the wavelength is $48.85 \mu$m [$49.85 \mu$m]. (a-b) For $\phi=0$ deg, (c-d) $\phi=135$ deg, and (e-f) $67.2$ deg. The inset in (f) zooms the gap region up to a maximum of 0.05 in the colorbar.}
\end{figure*}

 Figs.~\ref{fig:5_Perp_NF_Poynting}a-d [e-f] consist of maps related to the energy flow when $B$ is off [on] upon different azimuthal angles. The wavelengths coincide with that for the strongest SPhP in each case. The left column (a-c-e) shows the Poynting field $\mathbf S$ when $z=0$. Similarly, the right column (b-d-f) shows the z-component of $\nabla \times \mathbf S$. The white arrows are rescaled to visualize the maps easily. Interestingly,  Figs.~\ref{fig:5_Perp_NF_Poynting}a-b show an example of a repulsion state with zero spins when $\phi=0$, see Figs.~\ref{fig:3_Perp_B0T}a and b. Even though $\mathbf S$ aligns in a single direction, a resonant magnitude and a non-negligible curl appear near the surface of the particles. This resonance is due to the excitation of the SPhP. However, the contributions to the spin cancel out due to high symmetry evidenced by these maps and zero net spin results for the system. Figs.~\ref{fig:5_Perp_NF_Poynting}c-d show the attraction state with maximum positive spin when $\phi=135$ deg, see Figs.~\ref{fig:3_Perp_B0T}a and b. This time, two hot spots of maximum magnitude face each other, and a kind of saddle point emerges in the gap region between the particles, Fig.~\ref{fig:5_Perp_NF_Poynting}c. As a result, the values of the curl clearly show a rotational state for light as the field spots have "turbine-blade" shapes, Fig.~\ref{fig:5_Perp_NF_Poynting}d, explaining the net positive spin calculated for the system in Fig.~\ref{fig:3_Perp_B0T}b for $\phi=135$ deg. It is also evident from Fig.~\ref{fig:5_Perp_NF_Poynting}d that the two particles have the same spin, visually showing that the net spin is two times the spin of one particle. 
 Finally, we show the example when $B$ is on and $\phi=67.2$ deg, which coincides with the hot spot of maximum attraction at the resonance $49.85 \mu$m of the SPhP, see Fig.~\ref{fig:4_Perp_B1T}b. Notice in passing that this value for $\phi$ is close to the angle of the electric spots in Fig.~\ref{fig:4_Perp_B1T}d, namely, $\approx 65$ deg. Remarkably, Fig.~\ref{fig:5_Perp_NF_Poynting}e illustrates that the energy flow would make the particles spin counterclockwise. Moreover, it is also noteworthy the vortex that appears in the field region between the particles with clockwise orientation, resembling a "gear" mechanism which coordinates field and particles \cite{boriskina_plasmonics_2013,ekeroth_optical_2019-1,parker_optical_2020}. Consistently, the curl's map shows a structure similar to that in Fig.~\ref{fig:5_Perp_NF_Poynting}d but this time enhanced and possessing a structure in the gap region that contains the vortex indicated in Fig.~\ref{fig:5_Perp_NF_Poynting}e, see inset in Fig.~\ref{fig:5_Perp_NF_Poynting}f. The inset zooms this region so that a connection between the curl hot spots is appreciated.

\section{Conclusions}

By a simple dipolar model, this work explores the behavior of small nanoparticle dimers when magneto-optical materials like n-doped InSb and moderate magnetic fields are used. Two counter-propagating waves with equal circular polarization are used as illumination to simulate a simple optical trap with neither net gradient nor scattering forces. Our results show that the system can be thoroughly characterized by observing its mechanical inductions, provided these latter depend on the near field. Besides, we found no possibility of forming stable dimers when the dimer is aligned with the illumination since the inter-particle force only leads to repulsion. On the contrary, under "perpendicular" alignment, there are several ways to obtain stable dimers or inter-particle attraction, at least under this "static" model for which the particles' velocities and accelerations are not considered. As the results strongly depend on the magnetic field's presence, this study constitutes a novel background to build OM or photonic molecule nano factories and control their movements, or conversely, to study samples containing this class of multimers.
Finally, we give an original explanation for the appearance of particle spins based on the energy flows. This interpretation offers satisfactory results from the "scattering" forces produced by the interaction between the particles. Gradient forces were also investigated but showed no appreciable influence on the spins' appearances (results not shown in this work).

\section*{acknowledgments} 
	The author would like to thank A.~Garc\'ia-Mart\'in from IMN-CSIC and M.I. Marqués from Universidad Autónoma de Madrid for their valuable discussions during his postdoc in Spain.

\section*{Conflicts of Interest}

The authors declare that the research was conducted in the absence of any commercial or financial relationships that could be construed as a potential conflict of interest.

\bibliographystyle{iopart-num}

\end{document}